\documentclass{moriond}
\usepackage[nameinlink,capitalise,noabbrev]{cleveref}
\usepackage[caption=false]{subfig}
\usepackage{xspace}
\usepackage{amsbsy}
\usepackage{booktabs}

\def\Mbc{\ensuremath{M_{\mathrm{bc}}}\xspace}
\def\dE{\ensuremath{\Delta E}\xspace}
\def\lumi{\ensuremath{189\,\mathrm{fb^{-1}}}\xspace}
\def\ufours{\ensuremath{\Upsilon(4S)}\xspace}
\def\jpsi{\ensuremath{{J\mskip -3mu/\mskip -2mu\psi\mskip 2mu}}\xspace}

\def\gev{\ensuremath{\mathrm{\,Ge\kern -0.1em V}}\xspace}
\def\gevc{\ensuremath{{\mathrm{\,Ge\kern -0.1em V\!/}c}}\xspace}
\def\gevcc{\ensuremath{{\mathrm{\,Ge\kern -0.1em V\!/}c^2}}\xspace}
\def\invfb{\ensuremath{\mbox{\,fb}^{-1}}\xspace}
\def\epem{\ensuremath{e^+e^-}\xspace}
\def\mumu{\ensuremath{\mu^+\mu^-}\xspace}
\def\Bz{\ensuremath{B^0}\xspace}
\def\Bp{\ensuremath{B^+}\xspace}
\def\Kp{\ensuremath{K^+}\xspace}
\def\KS{\ensuremath{K^0_{\mathrm{S}}}\xspace}
\def\Kstar{\ensuremath{K^{*}(892)}\xspace}
\def\Kstarz{\ensuremath{K^{*0}(892)}\xspace}
\def\Kstarp{\ensuremath{K^{*+}(892)}\xspace}
\def\BB{\ensuremath{B\overline{B}}\xspace}
\def\Btag{\ensuremath{B_{\mathrm{tag}}}\xspace}
\newcommand\Tstrut{\rule{0pt}{2.6ex}}         
\newcommand\Bstrut{\rule[-0.9ex]{0pt}{0pt}}   

\newcommand{\Photo}{}

\begin{document}
\vspace*{4cm}
\title{Electroweak penguins and radiative B decays at Belle II}
\author{C.~Praz, on behalf of the Belle II collaboration}
\address{KEK, High Energy Accelerator Research Organization\\
Tsukuba, Japan}
\maketitle\abstracts{
Decays of $B$ mesons involving the transition of a $b$ quark into an $s$ quark are good probes of physics beyond the standard model.
Such decays are studied at the Belle II experiment, a detector located along the SuperKEKB electron-positron collider, and with data corresponding to an integrated luminosity of \lumi collected at the energy of the \ufours resonance.
The radiative decay of a $B$ meson into inclusive final states involving a strange hadron and a photon ($B \to X_s\gamma$) is studied by fully reconstructing the partner $B$ meson in a hadronic decay.
The photon-energy spectrum and the branching fraction of the inclusive $B \to X_s\gamma$ decay are reported.
In addition, the branching fraction of the decay of a $B$ meson into an excited $K$ meson and a pair of charged leptons ($B \to K^{\ast}(892)\ell^+\ell^-$, with $\ell^+\ell^-$ either an electron-positron or a muon-antimuon pair) is reported.
A control channel for the study of the charmless $B \to K\ell^+\ell^-$ decay is the decay $B \to J/\psi (\ell^+\ell^-) K$, for which a lepton-flavour universality ratio is reported.
All the reported measurements are in agreement with the world averages and the standard model predictions.
}
\section{Introduction}
\begin{figure}[b]
\centering
\subfloat[$b \to s\ell^+\ell^-$]{\includegraphics[width=0.325\linewidth]{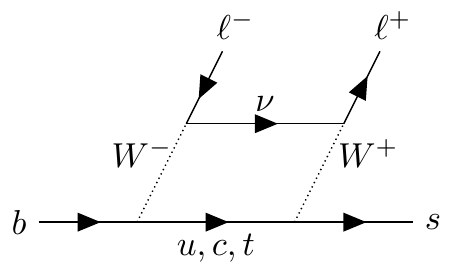}}
\subfloat[$b \to s\ell^+\ell^-$]{\includegraphics[width=0.325\linewidth]{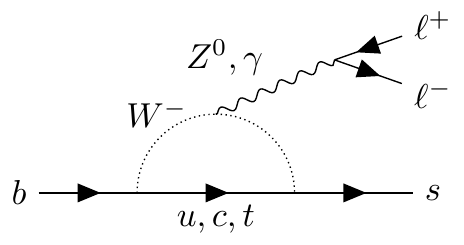}}
\subfloat[$b \to s\gamma$]{\includegraphics[width=0.325\linewidth]{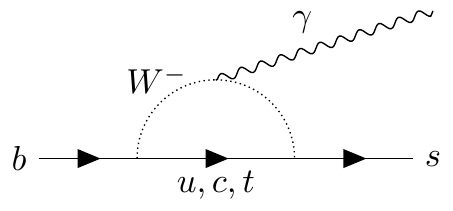}}
\caption{Lowest-order quark-level diagrams for the $b\to s \ell^+\ell^-$ (a)(b) and the $b \to s\gamma$ (c) transitions.}
\label{fig:feynman}
\end{figure}

The transition of a $b$ quark into an $s$ quark, which involves a flavour-changing neutral current, is an excellent probe of physics beyond the standard model (SM).
In the SM, flavour-changing neutral currents are forbidden at the tree level, meaning that the lowest-order Feynman diagrams contributing to a $b\to s$ transition are of the loop type or the box type, as illustrated in \cref{fig:feynman}.
These loop or box diagrams have a low probability of occurrence, implying that the decays of $B$ mesons based on a $b \to s$ transition are rare, with branching fractions typically ranging from $\mathcal{O}(10^{-7})$ to $\mathcal{O}(10^{-4})\,$ \cite{ParticleDataGroup:2022pth}.
In these decays, new particles or interactions may contribute to the Feynman diagrams at the same level as the SM particles and interactions, and affect the measured branching fractions and angular distributions of the final-state particles.

Decays of $B$ mesons are studied by the Belle II experiment~\cite{Abe:2010gxa}, which is located along the SuperKEKB collider~\cite{AKAI2018188} in Tsukuba, Japan.
SuperKEKB is colliding electrons and positrons at the energy of the \ufours resonance to produce pairs of $B$ mesons, and Belle II is designed to detect the products of the $B$-meson decays.
Since the beginning of the data collection in March 2019, Belle II has recorded an integrated luminosity of $362\invfb$ at the \ufours energy, which corresponds to $387\times10^6$ pairs of $B$ mesons, with an uncertainty of the order of 1\%.
The studies presented below are based on a partial data sample of $189\invfb$ collected between 2019 and 2021.

In these studies, two important variables are derived for a given $B$-meson candidate: the beam-constrained mass (\Mbc) and the energy difference (\dE).
They are defined as
\begin{equation}\label{eq:mbc_de}
    \Mbc \equiv \sqrt{\left(\frac{\sqrt{s}}{2}\right)^2 - p^{\ast 2}_B},
    \hspace{1cm}\Delta E \equiv E^\ast_B - \frac{\sqrt{s}}{2},
\end{equation}
where $\sqrt{s}$ is the available energy in the center-of-mass system (CMS), and $E^\ast_B$ and $p^{\ast}_B$ are the energy and momentum of the $B$-meson candidate in the CMS, respectively.

This report is organised as follows: 
\cref{sec:xsgamma} presents a measurement of the photon-energy spectrum and branching fraction of the inclusive $B \to X_s\gamma$ decay and is based on Ref.~\cite{Belle-II:2022hys},
\cref{sec:jpsik} presents the measurement of a lepton-flavour universality ratio for the $B \to J/\psi (\ell^+\ell^-) K$ decay, with $\ell=e,\mu$, and is based on Ref.~\cite{Belle-II:2022dbo},
and \cref{sec:kstarll} presents a measurement of the branching fraction of the $B \to K^{\ast}(892)\ell^+\ell^-$ decay and is based on Ref.~\cite{Belle-II:2022fky}.
\section{Measurement of the photon-energy spectrum in inclusive $B \to X_s\gamma$ decays}\label{sec:xsgamma}

This section reports a study of the inclusive $B \to X_s\gamma$ decay, where $X_s$ denotes any final state with a net strangeness.
In addition to its sensitivity to new physics for the reasons mentioned in introduction, this decay is also sensitive to the motion of the $b$ quark inside the $B$ meson~\cite{Bernlochner:2020jlt}.

The main experimental challenge is to suppress and subtract large background contributions coming from $\epem\to\BB$ and $\epem\to q\overline{q}$ events, where $q=u,d,c,s$ quarks.
The strategy for background suppression and signal selection relies on a technique called $B$-tagging, in which the partner $B$ meson (\Btag) in an $\epem\to\BB$ event is reconstructed to infer information about the signal decay.

The first step of the event selection is to reconstruct \Btag candidates in hadronic decays.
An algorithm called full event interpretation (FEI~\cite{Keck:2018lcd}) uses tracks and energy deposits in the Belle II calorimeter to reconstruct final-state particle candidates and combine them to form intermediate-state particle candidates.
The FEI combines final-state and intermediate-state particle candidates to build \Btag candidates in more than 30 hadronic modes.

The signal-photon candidate is reconstructed from an energy deposit in the Belle II calorimeter.
The energy of the signal-photon candidate in the signal $B$-meson frame, noted $E_\gamma^B$, is inferred from the kinematic properties of the \Btag candidate and the beam energy.
Only signal-photon candidates with $E_\gamma^B>1.4\gev$ are considered, and the highest-energy candidate is chosen in case of presence of multiple candidates in a single event.

Background arising from $\pi^0\to\gamma\gamma$ and $\eta\to\gamma\gamma$ decays is suppressed with a multivariate statistical-learning classifier that combines information from the signal-photon candidate and other photon candidates in the event.
Background due to $\epem\to q\overline{q}$ events is suppressed with another multivariate statistical-learning classifier trained in particular with variables that are functions of the distribution of the momenta in the event.

\begin{figure}[tb]
\centering
\includegraphics[width=0.495\linewidth]{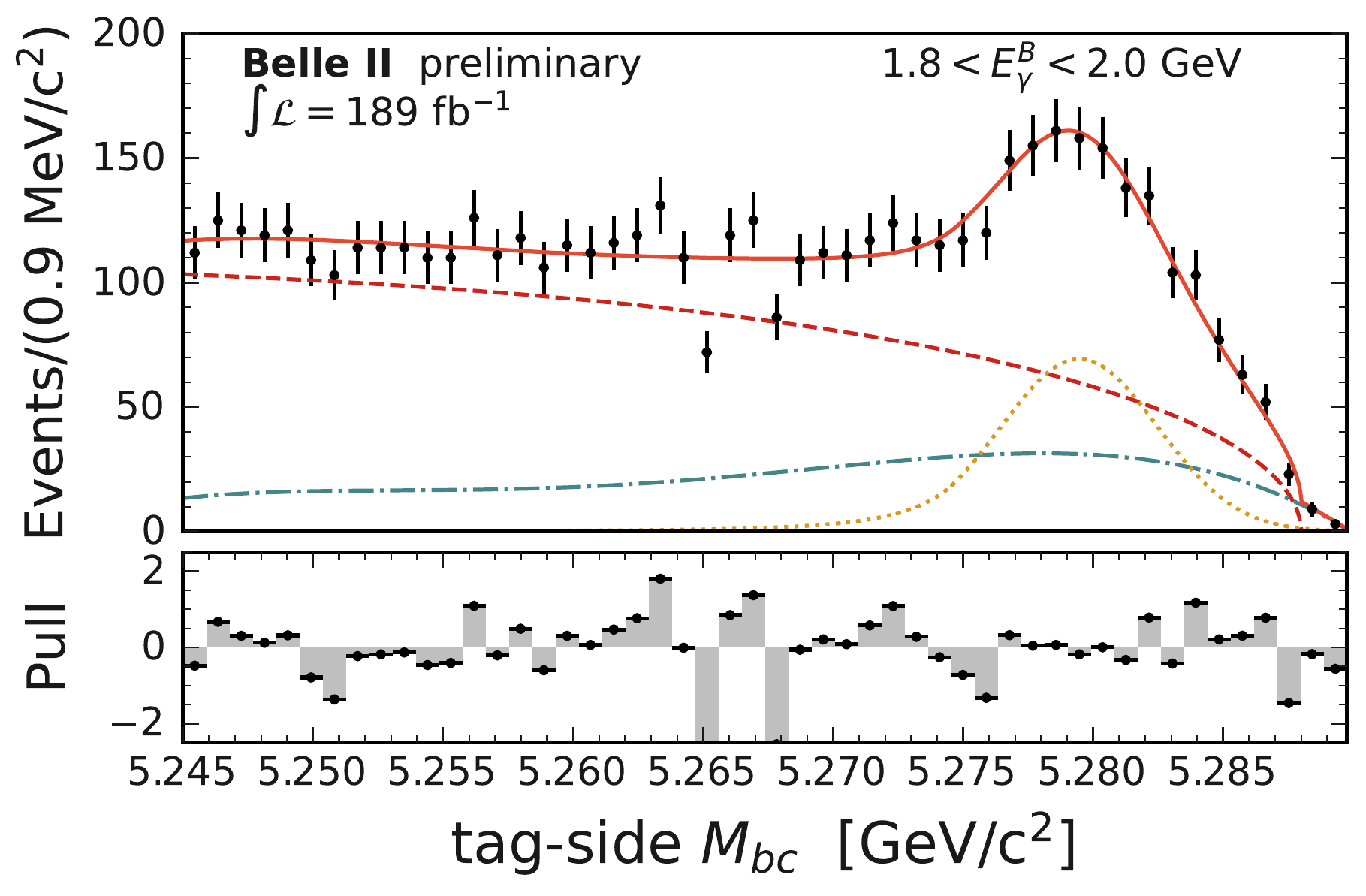}
\caption{Beam-constrained mass of the $B_{\mathrm{tag}}$ candidate in one interval of the signal-photon energy in data (black dots with error bars) and corresponding fit projection (solid red curve).
The three components of the fit are shown: correctly reconstructed $B_{\mathrm{tag}}$ candidates (orange dotted curve), mis-reconstructed $B_{\mathrm{tag}}$ candidates (blue dash-dotted curve), and background from $\epem\to q\overline{q}$ events, where $q=u,d,c,s$ quarks (red dashed curve).
Figure taken from Ref.~{\protect\cite{Belle-II:2022hys}}.
}
\label{fig:mbc_tag}
\end{figure}

The yield of $\epem\to\BB$ events with a correctly reconstructed $B_{\mathrm{tag}}$ candidate is determined from a fit to the \Mbc distribution of the \Btag candidate.
The fit is done simultaneously in multiple $E_\gamma^B$ intervals, one of which is shown in \cref{fig:mbc_tag}.

\begin{figure}[tb]
\centering
\subfloat[]{\includegraphics[width=0.495\linewidth]{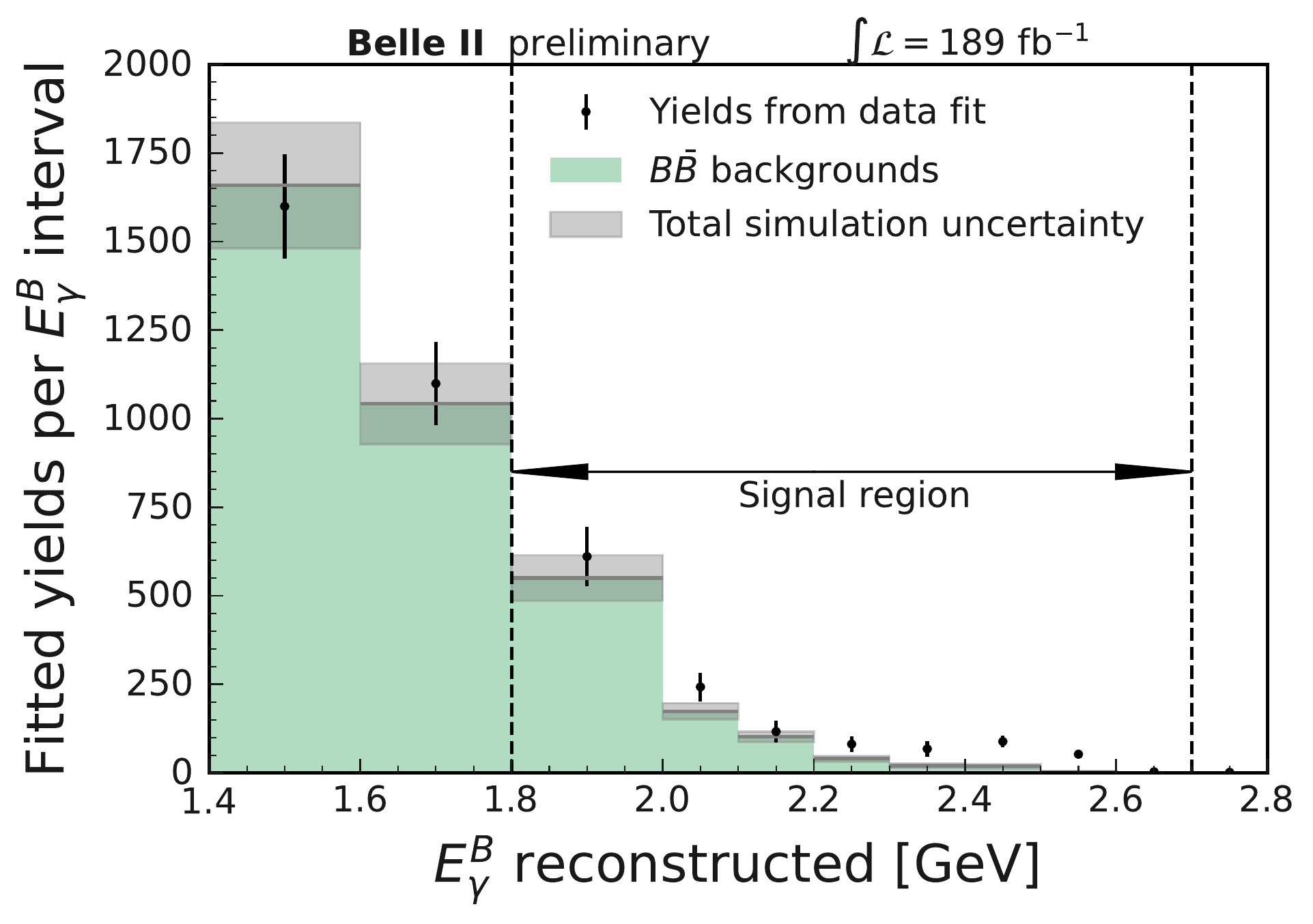}}
\subfloat[]{\includegraphics[width=0.495\linewidth]{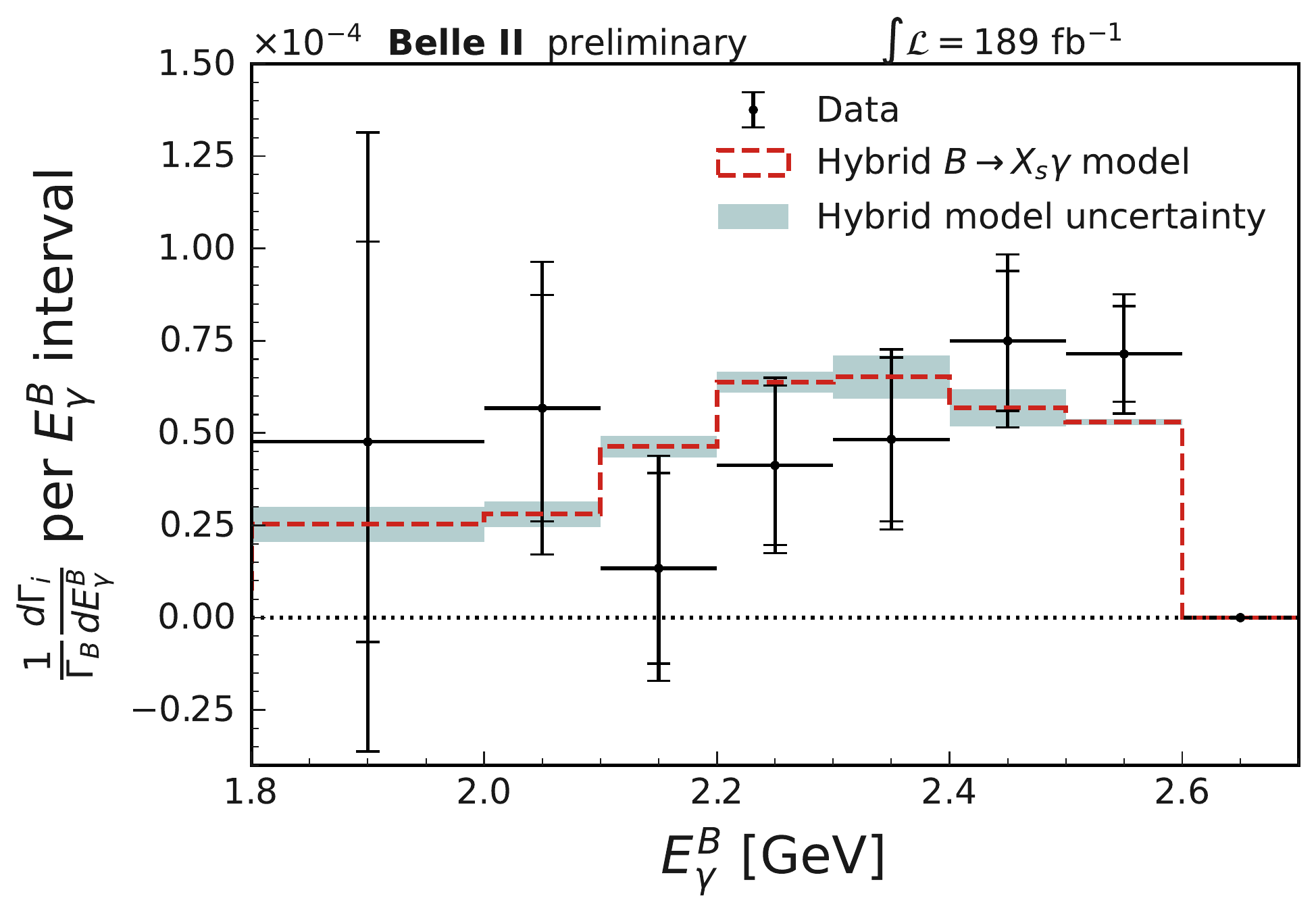}}
\caption{
Fitted yields of \BB events (a) and partial branching fraction of the inclusive $B\rightarrow X_s \gamma$ decay (b) in intervals of the photon energy in the $B_{\mathrm{sig}}$ frame.
In (a), the data points (black dots with error bars) correspond to the yields of \BB events with correctly reconstructed $B_{\mathrm{tag}}$ derived from the fit to data, and the histogram corresponds to yields determined from a fit to background-only simulated events.
In (b), the data points (black dots with error bars) are the measured branching fractions, with the inner (outer) error bars denoting the statistical (total) uncertainty, and the histogram shows the predictions of the signal model. 
Figure taken from Ref.~{\protect\cite{Belle-II:2022hys}}.}
\label{fig:egamma}
\end{figure}

Events with a correctly reconstructed $B_{\mathrm{tag}}$ candidate are either signal events or other, correctly tagged, $\epem\to\BB$ events.
The latter case represents a remaining background, whose yield in each $E_\gamma^B$ interval is determined from a fit to the \Mbc distribution of background-only simulated events.
Subtracting the yield of the remaining background (determined from the fit to background-only simulated events) from the total yield of events with a correctly reconstructed $B_{\mathrm{tag}}$ candidate (determined from the fit to data) allows to measure the branching fraction of the decay in intervals of $E_\gamma^B$ (\cref{fig:egamma}).

\cref{tab:xsgamma} shows the number of signal candidates after the background subtraction and the measured branching fraction for several $E_\gamma^B$ thresholds.
The systematic uncertainties are dominated by the background mis-modelling.
The result is compatible with the world average and competitive with the only other hadronic-tagging measurement from Ref.~\cite{BaBar:2007yhb} by the BaBar experiment with a similar sample size.

\begin{table}[tb]
\caption{Branching fraction of the inclusive $B\rightarrow X_s \gamma$ decay measured with the hadronic-tagging method by Belle II for two thresholds on the photon energy in the $B_{\mathrm{sig}}$ frame.
For each measured branching fraction, the first (second) quoted uncertainty denotes the statistical (systematic) uncertainty.
The integrated luminosity ($L$) of the dataset collected at the \ufours resonance and the number of signal candidates are also given.
}
\vspace{0.4cm}
\centering
\begin{tabular}{|lllllc|}
\hline
$E_\gamma^B$ threshold [$\gev$] & $\mathcal{B}(B\rightarrow X_s \gamma)$ [$10^{-4}$] & Experiment & L [$\invfb$] & Signal Yield & Ref.~\Tstrut\Bstrut\\
\hline
$1.8$ & $3.54 \pm 0.78 \pm 0.83$ & Belle II & 189 & $343\pm122$ & \cite{Belle-II:2022hys} \Tstrut\\
$2.0$ & $3.06 \pm 0.56 \pm 0.47$ & Belle II & 189 & $285\pm68$ & \cite{Belle-II:2022hys} \Bstrut\\
\hline
1.6 & $3.49\pm0.19$ & World average &  &  & \cite{ParticleDataGroup:2022pth} \Tstrut\Bstrut\\
\hline
\end{tabular}
\label{tab:xsgamma}
\end{table}
\section{Measurement of a lepton-flavour universality ratio in $B \to J/\psi K$ decays}\label{sec:jpsik}

The $B \to J/\psi (\ell^+\ell^-) K$ decay (with $\ell=e,\mu$) serves as a control channel for the study of the charmless $B \to K\ell^+\ell^-$ decay, since both decays share the same final-state particles.
The former decay is allowed at the tree-level in the SM, and has therefore a larger branching fraction than the latter.
This section reports a measurement of the lepton-flavour universality ratio $R_{K}{\left(J/\psi\right)}$, which is defined and measured as 
\begin{equation}\label{eq:luratio}
    R_{K}{\left(J/\psi\right)} \equiv 
    \frac{\mathcal{B}\left(B\to J/\psi(\mu^{+}\mu^{-})K\right)}{\mathcal{B}\left(B\to J/\psi(e^{+}e^{-})K\right)}=
    \frac{n_{\rm sig}^{J/\psi(\mu^{+}\mu^{-})K}}{n_{\rm sig}^{J/\psi(e^{+}e^{-})K}} \cdot
    \frac{\varepsilon^{J/\psi(e^{+}e^{-})K}}{\varepsilon^{J/\psi(\mu^{+}\mu^{-})K}},
\end{equation}
where $n_{\rm sig}^{J/\psi(\ell^{+}\ell^{-})K}$ and $\varepsilon^{J/\psi(\ell^{+}\ell^{-})K}$ are the number of signal candidates and the signal selection efficiency of the $B\to J/\psi(\ell^{+}\ell^{-})K$ mode, respectively.

Charged pion, kaon and lepton candidates are reconstructed from the information collected by the tracking subdetectors of Belle II.
A particle identification system (PID) combining the information of multiple subdetectors allows to suppress the background from mis-identified particle candidates.
Pairs of oppositely-charged lepton candidates are used to define \jpsi candidates, and pair of oppositely-charged pion candidates to define \KS candidates.
Requirements on the invariant mass of the \jpsi and \KS candidates suppress the background coming from wrong combinations. 
$B$-meson candidates are built by combining a \jpsi and a kaon candidate.

\begin{figure}[tb]
\centering
\subfloat[]{\includegraphics[width=0.495\linewidth]{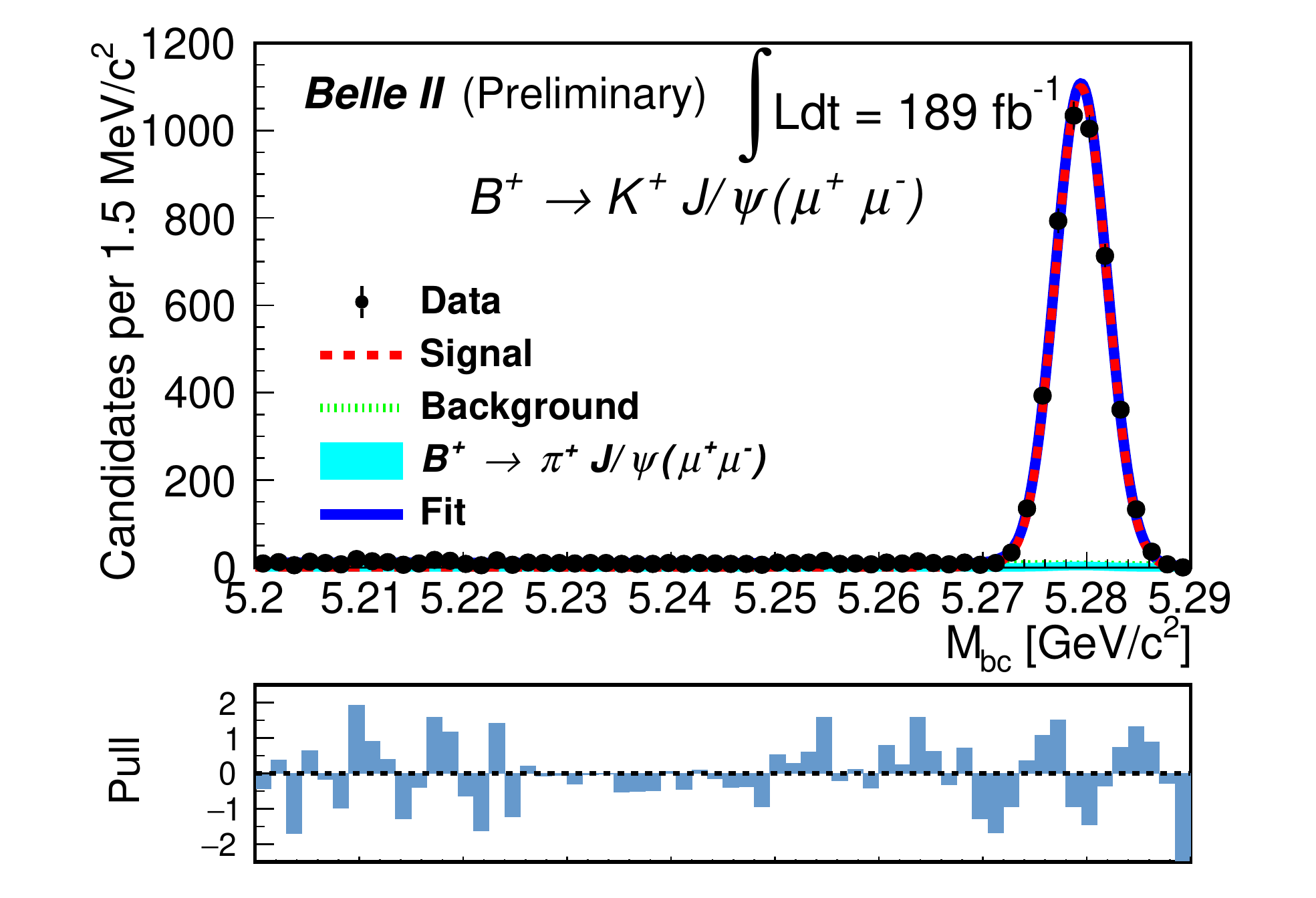}}
\subfloat[]{\includegraphics[width=0.495\linewidth]{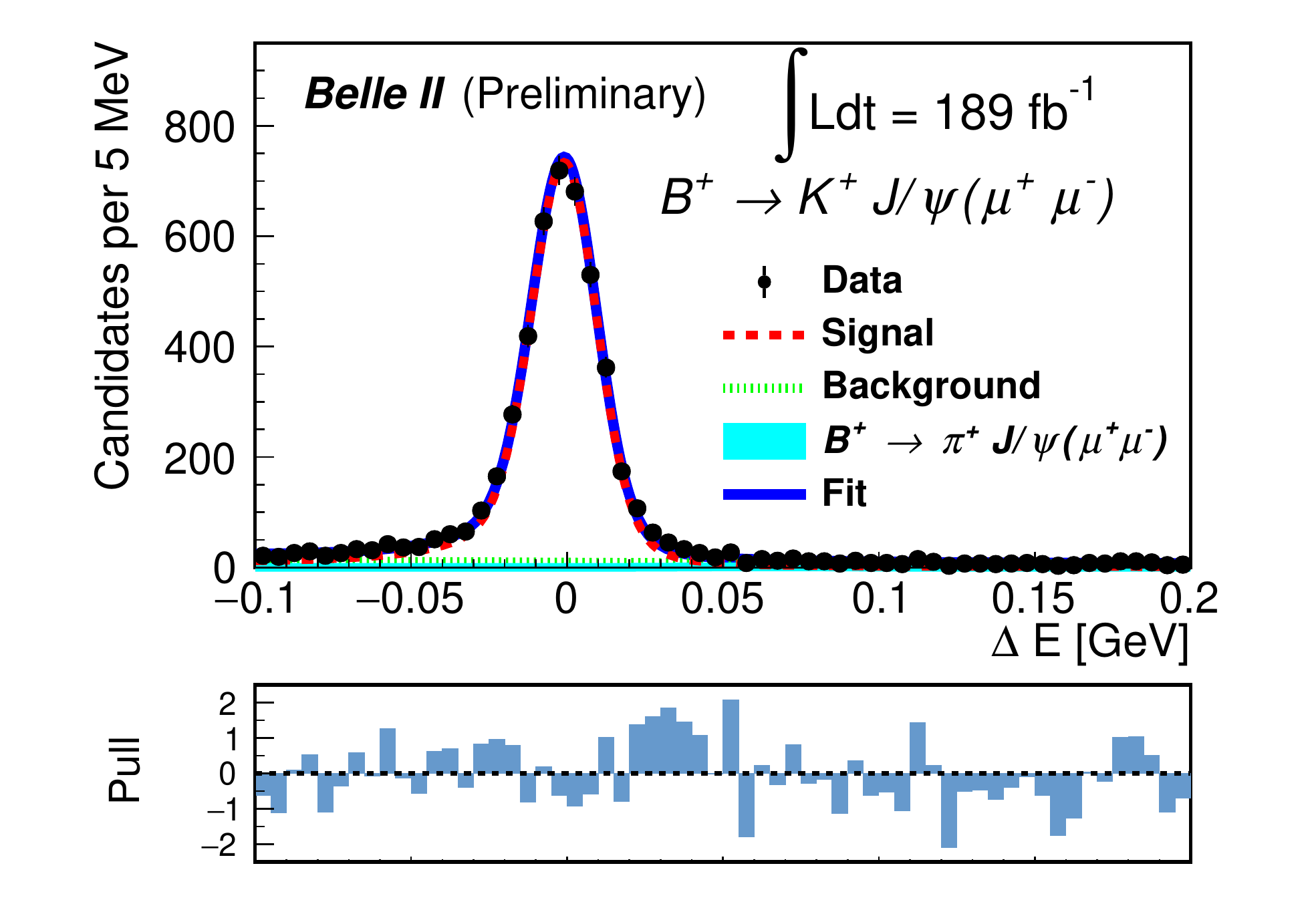}}
\caption{Beam-constrained mass (a) and energy difference (b) of $\Bp \to J/\psi(\mumu) \Kp$ candidates for data (black dots with error bars) and as predicted by the fit (blue curve).
The three components of the fit are shown: signal candidates (red dashed curve), background from $\Bp \to J/\psi(\mumu) \pi^ +$ (blue histogram), and remaining background (green dashed curve).
Figure taken from Ref.~{\protect\cite{Belle-II:2022dbo}}.}
\label{fig:jpsik}
\end{figure}

\cref{fig:jpsik} shows the yields of signal and background candidates as determined from a combined fit to \Mbc and \dE, defined in \cref{eq:mbc_de}, for the $\Bp \to J/\psi(\mumu) \Kp$ mode.
After selection, the sample is largely dominated by signal candidates, \Mbc is peaking at the mass of the $B$ meson, and \dE is peaking at zero.
For this mode, one important remaining background is coming from the $\Bp \to J/\psi(\mumu) \pi^ +$ decay, where the pion is mis-identified as a kaon.

The results are summarised in \cref{tab:jpsik}.
The measured lepton-flavour universality ratios are compatible with unity, and the uncertainty on each ratio is largely dominated by its statistical component.
On the other hand, the systematic uncertainty achieved by Belle II is better than the one published by Belle in Ref.~\cite{BELLE:2019xld}.

\begin{table}[tb]
\caption{Number of $\Bp\to J/\psi(\ell^+\ell^-)\Kp$ and $\Bz\to J/\psi(\ell^+\ell^-)\KS$ candidates, with $\ell=e,\mu$, and measured lepton-flavour universality ratio $R_{K}\left(J/\psi\right)$ defined in \cref{eq:luratio} and taken from Ref.~{\protect\cite{Belle-II:2022dbo}}.
For $R_{K}\left(J/\psi\right)$, the first (second) quoted uncertainty denotes the statistical (systematic) uncertainty.}
\vspace{0.4cm}
\centering
\begin{tabular}{|llll|}
\hline
Mode & $n_{\rm sig}^{J/\psi(\mumu)K}$ & $n_{\rm sig}^{J/\psi(\epem)K}$ & $R_{K}\left(J/\psi\right)$ \Tstrut\Bstrut\\
\hline
$\Bp\to J/\psi \Kp$ & $4578\pm62$ & $3706\pm62$ & $1.009 \pm 0.022 \pm 0.008$ \Tstrut\\
$\Bz\to J/\psi \KS$ & $1343\pm37$ & $1052\pm33$ & $1.042 \pm 0.042 \pm 0.008$ \Bstrut\\
\hline
\end{tabular}
\label{tab:jpsik}
\end{table}

\section{Measurement of the branching fraction of $B \to K^{\ast}(892)\ell^+\ell^-$  decays}\label{sec:kstarll}

This section reports a measurement of the branching fractions of the $B \to K^{\ast}(892)\ell^+\ell^-$ decays, with $\ell=e,\mu$.
These branching fractions are very small, of the order of $10^{-6}$, making their determination challenging due to the relatively large level of background.


As in \cref{sec:jpsik}, charged pion, kaon and lepton candidates are reconstructed from the information collected by the tracking subdetectors of Belle II and selected using their PID information.
Neutral pion candidates are reconstructed from pairs of photon candidates, and $\KS$ candidates from pairs of oppositely-charged pion candidates.
Excited kaon candidates are reconstructed in the $\Kstarz\to\Kp\pi^-$, $\Kstarp\to\Kp\pi^0$ and $\Kstarp\to\KS\pi^+$ modes, and $B$-meson candidates are built by combining a \Kstar candidate and a pair of oppositely-charged lepton candidates.

Background coming from $B\to\jpsi\Kstar$, $B\to\psi(2S)\Kstar$ and $B\to\gamma\Kstar$ decays is suppressed by vetoing regions of the di-lepton invariant mass.
In addition, a multivariate statistical-learning classifier is trained on simulated events to suppress the background coming from non-resonant $\epem\to q\overline{q}$ events, where $q=u,d,c,s$ quarks.

\begin{figure}[tb]
\centering
\subfloat[]{\includegraphics[width=0.495\linewidth]{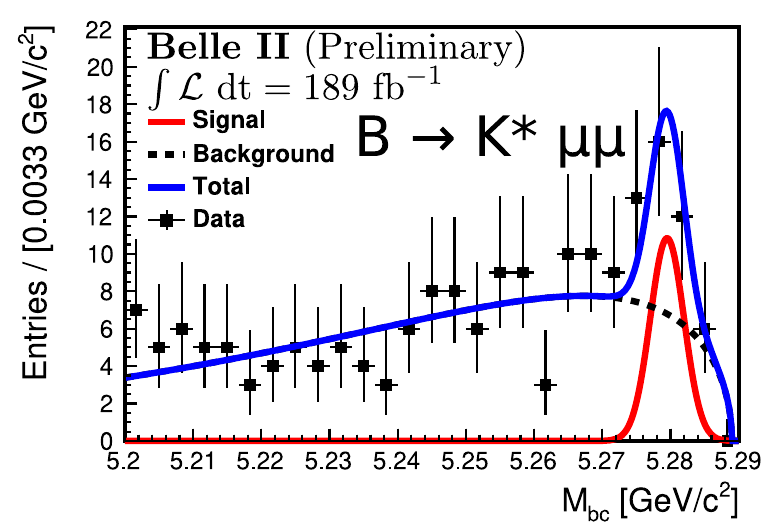}}
\subfloat[]{\includegraphics[width=0.495\linewidth]{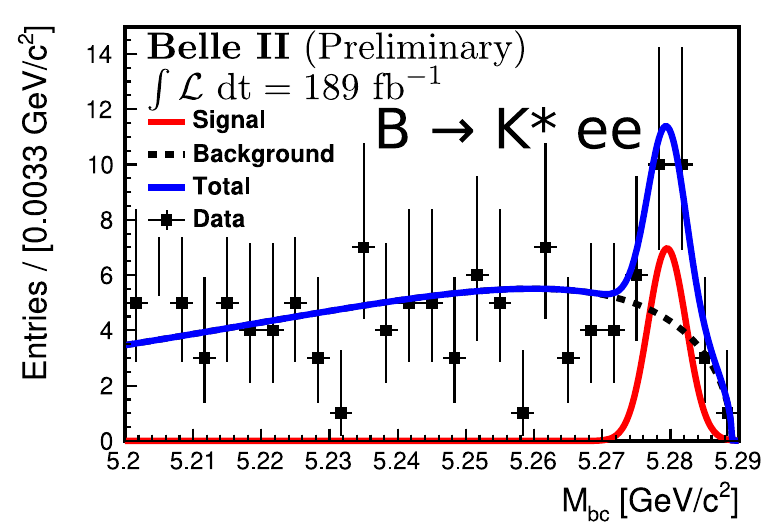}}
\caption{Beam-constrained mass of $B \to K^{\ast}(892)\mu^+\mu^-$ candidates (a) and $B \to K^{\ast}(892)e^+e^-$ candidates (b) for data (black dots with error bars) and as predicted by the fit (blue curve).
The two components of the fit are shown: signal candidates (red curve), and background (black dashed curve).
Figure taken from Ref.~{\protect\cite{Belle-II:2022fky}}.}
\label{fig:kstarll}
\end{figure}

Similarly to what is done in \cref{sec:jpsik}, the signal and background yields are determined from a combined fit to \Mbc and \dE, defined in \cref{eq:mbc_de}.
\cref{fig:kstarll} shows the result of the fit in \Mbc for the electron and muon modes.
The remaining background mainly comes from mis-reconstructed $\epem\to\BB$ events.

The number of signal candidates and measured branching fractions are presented in \cref{tab:kstarll}.
The results are compatible with the world averages, and the uncertainty on each branching fraction is largely dominated by its statistical component.
A comparison between the $B\to\Kstar\mu^+\mu^-$ mode and the $B\to\Kstar e^+e^-$ mode shows that Belle II achieves a good and similar identification of muons and electrons.

\begin{table}[tb]
\caption{Number of $B\to\Kstar\mu^+\mu^-$ and $B\to\Kstar e^+e^-$ candidates and measured branching fractions taken from Ref.~{\protect\cite{Belle-II:2022fky}}.
For each measured branching fraction, the first (second) quoted uncertainty denotes the statistical (systematic) uncertainty.
The world average values are taken from Ref.~{\protect\cite{ParticleDataGroup:2022pth}}.}
\vspace{0.4cm}
\centering
\begin{tabular}{|llll|}
\hline
Observable & Signal Yield & Measured value [$10^{-6}$] & World average [$10^{-6}$] \Tstrut\Bstrut\\
\hline
$\mathcal{B}(B\to\Kstar\mu^+\mu^-)$ & $22\pm6$ & $1.19 \pm 0.31^{+0.08}_{-0.07}$ & $1.06\pm0.09$ \Tstrut\\
$\mathcal{B}(B\to\Kstar e^+e^-)$ & $18\pm6$ & $1.42 \pm 0.48\pm 0.09$ & $1.19\pm0.20$ \Bstrut\\
\hline
\end{tabular}
\label{tab:kstarll}
\end{table}

\section{Summary}
The result presented in \cref{sec:xsgamma} and detailed in Ref.~\cite{Belle-II:2022hys} is the first measurement of the inclusive $b\to s\gamma$ transition with the hadronic-tagging method by the Belle II experiment.
This measurement is competitive with the only other hadronic-tagging measurement from Ref.~\cite{BaBar:2007yhb} by the BaBar experiment.
The results of the studies of $B \to J/\psi(\ell^+\ell^-) K$ and $B \to K^{\ast}(892)\ell^+\ell^-$ decays, presented in \cref{sec:jpsik,sec:kstarll}, and detailed in Refs.~\cite{Belle-II:2022dbo,Belle-II:2022fky}, are in agreement with the world averages; these results show a good and similar identification of electrons and muons by the Belle II experiment and prepare the ground for precision measurements of rare decays when more data will be collected.
\section*{References}

\end{document}